\documentclass[letterpaper,11pt]{article}
\usepackage[utf8]{inputenc}
\usepackage[margin=1in]{geometry}
\usepackage[T1]{fontenc}
\usepackage[english]{babel}
\usepackage{lmodern}
\usepackage{amsmath,amssymb,amsthm,mathtools}
\usepackage{microtype,booktabs,array,enumitem}
\usepackage{needspace}
\usepackage[dvipsnames]{xcolor}
\usepackage{xurl}
\usepackage[colorlinks=true,pdfpagemode=UseNone,urlcolor=RoyalBlue,linkcolor=RoyalBlue,citecolor=OliveGreen,pdfstartview=FitH]{hyperref}
\usepackage[nameinlink,capitalize,noabbrev]{cleveref}
\hypersetup{pdftitle={Short Point Lists for CNFs and Stronger Constant-Depth Circuit Hardness},pdfauthor={Xin Li and Yan Zhong}}
\setlist{itemsep=3pt,topsep=5pt}
\numberwithin{equation}{section}
\hypersetup{bookmarksnumbered=true}
\newtheorem{theorem}{Theorem}[section]
\newtheorem{lemma}[theorem]{Lemma}
\newtheorem{proposition}[theorem]{Proposition}
\newtheorem{corollary}[theorem]{Corollary}
\theoremstyle{definition}

\theoremstyle{remark}

\theoremstyle{plain}
\newcommand{\proj}{\pi}
\newcommand{\F}{\mathsf{F}}
\newcommand{\E}{\mathbf{E}}
\newcommand{\Prb}{\mathbf{Pr}}
\newcommand{\bits}{\{0,1\}}
\newcommand{\1}{\mathbf 1}
\newcommand{\AC}{\mathsf{AC}^{0}}
\newcommand{\cl}{\operatorname{cl}}
\newcommand{\Mass}{\operatorname{Mass}}
\newcommand{\Corr}{\operatorname{Corr}}
\newcommand{\Tr}{\operatorname{Tr}}
\newcommand{\inv}{\operatorname{inv}_{0}}

\newcommand{\ind}{\mathbf 1}
\newcommand{\eps}{\varepsilon}
\newcommand{\abs}[1]{\left|#1\right|}
\newcommand{\ceil}[1]{\left\lceil#1\right\rceil}
\newcommand{\floor}[1]{\left\lfloor#1\right\rfloor}

\title{Short Point Lists for CNFs and\\Stronger Constant-Depth Circuit Hardness}
\author{Xin Li\thanks{Department of Computer Science, Johns Hopkins University. Emails: \href{mailto:lixints@cs.jhu.edu}{\texttt{lixints@cs.jhu.edu}},
\href{mailto:yzhong36@jhu.edu}{\texttt{yzhong36@jhu.edu}}.}
\and
Yan Zhong\footnotemark[1]}
\date{October 18, 2026}
\begin{document}
\pagenumbering{gobble}
\hypersetup{pageanchor=false}
\maketitle
\enlargethispage{\baselineskip}
\begin{abstract}
Proving lower bounds of an explicit function for the class of constant depth circuits with unbounded fan-in $\mathsf{AND}$ and $\mathsf{OR}$ gates is a central question in circuit complexity. The celebrated result of H\aa stad (1989) gives $2^{\Omega(n^{1/(d-1)})}$ lower bounds for the parity and majority functions, which have been asymptotically the best known bounds for decades. 

Recently, OpenAI (2026) improved these bounds and gave an explicit function in $\mathsf{P}$ that requires $2^{\omega(\sqrt{n})}$ size $\mathsf{OR}$–$\mathsf{AND}$–$\mathsf{OR}$ circuits. Soon after, Chukhin (2026) improved OpenAI's result and obtained a lower bound of 
$2^{\Omega_d((n\log n/\log\log n)^{1/(d-1)})}$ against depth-$d$ circuits for every fixed $d \geq 3$. In this paper, we further improve Chukhin's result and prove a worst-case size lower bound of $2^{\Omega_d((n\log n)^{1/(d-1)})}$ for three explicit functions: inverse trace over binary fields, the Goldreich--Tal trilinear function, and the polynomial evaluator of OpenAI with its original parameters. For inverse trace, we also prove a correlation bound under the uniform distribution: every depth-at-most-$d$ circuit of size at most $S$, where $n\le S\le2^{c_dn^{1/(d-1)}}$, has correlation at most $2^{-\eta_d n\log\log S/(\log S)^{d-1}}$ with the inverse trace function. Here $c_d,\eta_d>0$ are constants that depend only on the depth $d$. This improves on the classical parity correlation bound given by H\aa stad (2014).

Our worst case lower bounds are obtained by combining the techniques in previous works with a novel short-description decomposition of bounded-width CNFs, and the correlation bound additionally uses common partial decision trees. We also construct exact nonuniform AND/OR circuits for inverse trace of size \(2^{O(n^{2/3})}\) at depth eight and \(2^{O(\sqrt n\log n)}\) at depth nine. More generally, for every fixed integer \(h\ge2\), the construction gives depth-\((4h+1)\) circuits of size \(2^{O_h(n^{1/h}\log n)}\). The upper bounds follow by implementing a finite-field evaluation and interpolation construction using circuits that enumerate intermediate states.

\end{abstract}
\clearpage
\begingroup
\tableofcontents
\endgroup
\clearpage
\pagenumbering{arabic}
\hypersetup{pageanchor=true}

\section{Introduction}
\label{sec:intro}
Circuit complexity is one of the most fundamental and well studied areas in theoretical computer science. Once viewed as a promising approach towards proving $\mathsf{NP} \neq \mathsf{P}$, the ultimate goal is to find some function in the class $\mathsf{NP}$ which requires super-polynomial circuit size. Unfortunately, the results remain far from satisfactory. For general fan-in-two Boolean circuits over the full binary basis, Li and Yang~\cite{LY22} proved an explicit size lower bound of $3.1n-o(n)$, and a recent preprint of Mishra and Santhanam~\cite{MS26} reports an improvement to $3.5n-o(n)$, despite the fact that a random Boolean function requires exponential circuit size with high probability. On the other hand, we do have exponential lower bounds for restricted circuit classes, and one of the most extensively studied cases is the class of $\mathsf{AC^0}$, which captures constant depth circuits with $\mathsf{AND}$ and $\mathsf{OR}$ gates of unbounded fan-in. 

Here, the celebrated result of H\aa stad \cite{Hastad89} (see also \cite{ajtai1983sigma11, furst1984parity, yao1985separating, razborov1987lower, smolensky1987algebraic}) shows that the parity and majority functions both require $2^{\Omega(n^{1/(d-1)})}$ size depth-$d$ $\mathsf{AC^0}$ circuits. Subsequent work has refined these bounds and developed alternative proofs~\cite{HJP95, GRSS23, Kor26, KR26}. For parity, this is well known to be optimal, i.e., there exists a size $2^{O(n^{1/(d-1)})}$ depth-$d$ $\mathsf{AC^0}$ circuit that computes parity. For majority, this was not known until the recent result of Lecomte and Ramakrishnan \cite{lecomte2026optimal}, which constructed depth-$d$ $\mathsf{AC^0}$ circuits of size $2^{O(n^{1/(d-1)})}$ for every symmetric function. Hence, the bound is also optimal for majority, and in particular, no symmetric function can break this bound.  

Breaking this barrier, and proving stronger lower bounds for $\mathsf{AC^0}$ circuits has since been one of the central open problems in circuit complexity. This question is important both in its own right and because of its connections to general circuit lower bounds. For example, Valiant \cite{Val77, Val83} showed that any function computed by a linear size Boolean circuit of fan-in two and logarithmic depth can also be computed by a depth-three $\mathsf{AC^0}$ circuit of size $2^{O(n/ \log \log n)}$, so an explicit function with a depth-three $\mathsf{AC^0}$ lower bound of $2^{\omega(n/ \log \log n)}$ would also give the first superlinear size lower bound for logarithmic depth general Boolean circuits. We refer the reader to \cite{Vio09, GKW21} for further discussions of the connections between $\mathsf{AC^0}$ lower bounds and general circuit lower bounds.

Four decades after H\aa stad's result, a recent paper by OpenAI~\cite{OpenAI26} breaks the depth three barrier, and gives an explicit function that requires $\mathsf{OR}$–$\mathsf{AND}$–$\mathsf{OR}$ circuits of size $2^{\omega(\sqrt n)}$. As observed by Chukhin \cite{Chukhin26}, their proof implicitly gives the quantitative bound of $2^{\Omega(\sqrt {n \log \log n /\log  \log  \log n})}$. Subsequently,  Chukhin~\cite{Chukhin26} improved the result and obtained a lower bound of 
\begin{equation}
2^{\Omega_d((n\log n/\log\log n)^{1/(d-1)})}
\label{eq:previous}
\end{equation}
at every fixed depth $d\ge3$ for a particularly simple trilinear function.

In this paper, we further improve Chukhin's result and obtain stronger lower bounds.

\subsection{Results}
Without loss of generality, we consider unbounded-fan-in AND/OR circuits with negations only at the inputs. This at most doubles the size of the $\mathsf{AC^0}$ circuit. Size counts all AND and OR gates, including bottom-layer gates; input negations are free. Depth is the largest number of gates on an input-to-output path. Arbitrary fan-out is allowed, so subcircuits may be shared. We obtain both worst case and average case hardness results. 
For average case hardness, we define the correlation of two Boolean functions $F,C$ on $n$ bits to be
\[
\Corr(F,C)=\abs{\E_x(-1)^{F(x)+C(x)}}.
\]
Thus the absolute deviation of the agreement probability from $1/2$ is $\Corr(F,C)/2$. All logarithms are base two unless written $\ln$.

\paragraph{Worst-case hardness.}
For a fixed binary basis of $E=\F_{2^N}$, let
\[
I_N(x)=\Tr_{E/\F_2}(\inv(x)),\qquad
\inv(x)=\begin{cases}x^{-1},&x\ne0,\\0,&x=0.\end{cases}
\]
Thus $I_N$ is the trace component of the inverse map, with inversion extended by $\inv(0)=0$. At $N=2^j$ we use an explicit quadratic tower; at other lengths we pad from the largest such $N$. Denote the resulting family by $F_n$. This family is fixed independently of the circuit depth and size.

\Needspace{15\baselineskip}
\begin{theorem}[Worst-case lower bounds]
\label{thm:worst}
For every fixed $d\ge3$, each of the following polynomial-time Boolean families requires depth-at-most-$d$ circuits of size
\[
2^{\Omega_d((n\log n)^{1/(d-1)})},
\]
where $n$ counts all input bits:
\begin{enumerate}
\item the padded inverse-trace family $F_n$;
\item the Goldreich--Tal trilinear function~\cite{GT18}
\[
T_a(x,y,z)=\sum_{0\le i,j<a}z_{i+j}x_i y_j\pmod2,
\qquad n=4a-1;
\]
\item the polynomial evaluator of OpenAI~\cite{OpenAI26}, defined explicitly in \Cref{sec:evaluator}, with its original hash dimension and polynomial degree.
\end{enumerate}
\end{theorem}

The coefficient input $z$ in the trilinear function is part of the input. Likewise the evaluator's hash, modulus, and polynomial coefficients are counted. Hard parameter settings can be selected nonconstructively in a lower-bound proof without changing the uniform algorithm computing the full function. At depth three the lower bound is $2^{\Omega(\sqrt{n\log n})}$.

\paragraph{Average-case hardness.}
The inverse-trace family has an additional property that permits control of arbitrary circuit correlation, uniformly from polynomial size to a subexponential size range.

\begin{theorem}[Correlation under the uniform distribution]
\label{thm:average}
For every fixed $d\ge3$ there are constants $c_d,\eta_d>0$ such that, for every sufficiently large $n$, every depth-at-most-$d$ circuit $C$ of size at most $S$ satisfies
\begin{equation}
n\le S\le 2^{c_dn^{1/(d-1)}}
\quad\Longrightarrow\quad
\Corr(F_n,C)\le
2^{-\eta_d n\log\log S/(\log S)^{d-1}}.
\label{eq:main-average}
\end{equation}
\end{theorem}

This improves the classical correlation benchmark of parity in \cite{Hastad14} by a factor of $\log \log S$ in the exponent. For $S=n^B$, with $B\ge1$ fixed, the exponent in Eqn.~\eqref{eq:main-average} is $\Omega_{d,B}(n\log\log n/(\log n)^{d-1})$. For $S=2^{n^\alpha}$ and fixed $0<\alpha<1/(d-1)$, it is $\Omega_{d,\alpha}(n^{1-\alpha(d-1)}\log n)$. For example, depth-three circuits of size $2^{n^{2/5}}$ have correlation at most $2^{-\Omega(n^{1/5}\log n)}$.

Tal's Fourier-tail theorem~\cite{Tal17}, together with the uniform bound on the Fourier coefficients of inverse trace, also gives the comparison bound $2^{-\Omega_d(n/(\log S)^{d-1})}$ for inverse trace. Our bound gains a factor $\log\log S$ in this exponent; the comparison is proved in \Cref{app:fourier}. This comparison concerns the bound obtained from Tal's theorem; we do not claim to subsume all previous correlation bounds. Also, the size range in \Cref{thm:average} is smaller than the worst-case threshold in \Cref{thm:worst}. The latter does not follow by substituting its threshold into Eqn.~\eqref{eq:main-average}.

\paragraph{Explicit upper bounds for inverse trace.}
The same inverse-trace family also admits subexponential-size circuits at sufficiently large fixed depths. Write $\operatorname{size}_d(G)$ for the minimum number of AND/OR gates in a depth-at-most-$d$ circuit computing $G$, with free input negations and arbitrary sharing. For a triple $\mathbf h=(h_1,h_2,h_3)$ of positive integers, put
\[
\Phi_{\mathbf h}(N)=
N^{1/h_1}(\log N)^{1-1/h_1}
+N^{1/h_2}\log N+N^{2/h_3}.
\]
Let $U_d(N)$ be the minimum of $N$ and these quantities over all such triples with $h_1+h_2+h_3+1\le d$; if $d=2,3$, set $U_d(N)=N$.

\begin{theorem}[Nonuniform upper bounds]
\label{thm:upper}
For every fixed integer $d\ge2$, inverse trace $I_N$ in any fixed binary basis has circuits satisfying
\begin{equation}
\operatorname{size}_d(I_N)\le2^{O_d(U_d(N))}.
\label{eq:main-upper}
\end{equation}
In particular,
\begin{equation}
\operatorname{size}_8(I_N)\le2^{O(N^{2/3})},\qquad
\operatorname{size}_9(I_N)\le2^{O(\sqrt N\log N)}.
\label{eq:upper-examples}
\end{equation}
More generally, for every fixed integer $h\ge1$,
\begin{equation}
\operatorname{size}_{4h+1}(I_N)\le
2^{O_h(\min\{N,N^{1/h}\log N\})}.
\label{eq:upper-symmetric}
\end{equation}
The same estimates hold when wires are counted, and their constants are independent of the chosen basis. They also hold for the padded family $F_n$ with $n$ in place of $N$.
\end{theorem}

Together with \Cref{thm:worst}, this places the exact circuit complexity of one fixed explicit family between
\begin{equation}
2^{\Omega_d((n\log n)^{1/(d-1)})}
\le\operatorname{size}_d(F_n)
\le2^{O_d(U_d(n))},\qquad d\ge3.
\label{eq:inverse-sandwich}
\end{equation}
The upper bound is subexponential at every fixed depth $d\ge8$. There remains a gap with the lower bound; at depths three and four, this construction gives only the full-DNF exponential estimate. We do not claim that the upper bounds are optimal or that depth eight is the first depth admitting subexponential circuits.

The proof uses the evaluation and interpolation approach to finite-field arithmetic of Healy and Viola~\cite{HV05}. We implement three stages directly at constant depth: linear evaluation followed by logarithm tables, multiplication of the evaluated factors, and a final parity. Each stage can be viewed as a computation with few state bits, and enumerating intermediate states gives an adjustable depth--size tradeoff. Choosing the gate orientations at the two stage boundaries saves two layers. \Cref{sec:upper} gives the construction and its depth accounting. All final gates are ordinary AND/OR gates, with negations only at the inputs.

\subsection{Proof Overview} 
We prove our lower bounds by establishing the following main technical theorem.
\paragraph{The structural theorem.}
Given a nonempty list $A$ of points of $\bits^m$, we define its width-$b$ CNF closure to be
\begin{equation}
\cl_b(A)=\{x:\ x|_U\in\{a|_U:a\in A\}\text{ for every }U\subseteq[m],\ |U|\le b\}.
\label{eq:closure-intro}
\end{equation}
In other words, a string $x$ belongs to $\cl_b(A)$ exactly when \emph{every} local pattern
of size at most $b$ exhibited by $x$ is exhibited by at least one
member of $A$.

We first note that $\cl_b(A)$ is the set of satisfying assignments of a width $b$ CNF. Indeed, for a subset $I\subseteq[m]$ define the projection $\pi_I: \pi_I(x)=x|_I$ and $\proj_I(A)=\{a|_I:a\in A\}$. Then for any $z\in\bits^I$ such that $z \notin \proj_I(A)$, any $x \in \cl_b(A)$ must satisfy that $\proj_I(x) \neq z$. Hence, let 
\begin{equation}\label{eq:forbidding-clause}
 C_{I,z}(x)=\bigvee_{i\in I}(x_i\ne z_i),
\end{equation}
then 
\begin{equation}
\boxed{\quad
\1_{\cl_b(A)}(x)
 =\bigwedge_{\substack{I\subseteq[m],\ |I|\le b\\
                        z\in\bits^I\setminus\proj_I(A)}} C_{I,z}(x).
\quad}
\end{equation}

It is also not hard to see that $\cl_b(A)$ is the \emph{smallest satisfying assignment set} of any width-$b$ CNF that contains every string in $A$. In this sense, we can use a list $A$ to describe a CNF, without listing its clauses. This is particularly useful because the CNF can have a large number of clauses.

\begin{theorem}[Point-list decomposition, informal form]
\label{thm:partition-informal}
For every fixed $\eps>0$, every width-$b$ CNF on $m\ge1$ variables is an exact disjoint union of at most $2^{\eps m}$ sets $\cl_b(A_i)$, where
\[
|A_i|\le 2^{O_\eps(b)}\log(m+1).
\]
No bound on the number of clauses in a piece, or on the time to find the decomposition, is asserted.
\end{theorem}

On a high level, our proof of lower bounds follows the same framework as in ~\cite{Chukhin26} and ~\cite{OpenAI26}, by first reducing the problem to bounding correlations with width-$b$ CNFs for an appropriate choice of $b$. The key step in the analysis is to make sure that restrictions of the hard function have small correlation with \emph{all} width-$b$ CNFs. However, the number of such CNFs is very large, hence ~\cite{Chukhin26} uses the CNF sparsification results developed by Impagliazzo, Paturi, and Zane~\cite{IPZ01}, and refined by Calabro, Impagliazzo, and Paturi~\cite{CIP06} to reduce the case to CNFs with few clauses. The number of such CNFs is much smaller than general width-$b$ CNFs, and therefore a union bound suffices. However, this approach still results in $b^{O_\varepsilon(b)}m$ clauses, which only permits $b=\Theta(\log n/\log\log n)$ as in ~\cite{Chukhin26}.

Here, we instead use our point-list decomposition, so that the problem is reduced to bounding the correlation with all width-$b$ CNFs that are described by a short list. The counting of such lists pays a better price in the union bound, replacing the \(b^{O_\varepsilon(b)}\) dependence from clause sparsification by \(2^{O_\varepsilon(b)}\), thus allowing $b=\Theta(\log n)$ even when the pieces have many clauses. Consequently, our theorem does not resolve the algorithmic CNF sparsification problem asking for pieces with only $2^{O(b)}m$ clauses.

\paragraph{Why short point lists improve the exponent.}
Let $\kappa_b(g)=\max_H|\E[gH]|$, where $H$ ranges over all width-$b$ CNFs and $g$ is a sign function. The maximum includes the constant-one CNF. Here $H$ is a $\{0,1\}$-valued indicator: $\kappa_b$ measures correlation with CNF indicators, whereas $\Corr$ uses the $\{-1,1\}$ encodings of both Boolean functions. A decomposition into $2^{\eps m}$ pieces gives
\[
\kappa_b(g)\le2^{\eps m}\max_{|A|\le L}|\E[g\ind_{\cl_b(A)}]|.
\]
There are fewer than $2^{mL+1}$ lists of length at most $L$ when $m\ge1$. The simultaneous moment argument therefore pays $O(mL)$ description bits. With our theorem, $L=2^{O_\eps(b)}\log(m+1)$; hence widths $b$ proportional to $\log n$ remain affordable.

Here is the parameter calculation behind the worst-case result. Write $\ell=\log S$. Standard depth reduction leaves a depth-three circuit with bottom fan-in $k=\Theta_d(\ell)$ and survival probability $q=\Theta_d(\ell^{-(d-3)})$. A second restriction of survival $p=b/(64k)$ transfers width $k$ to width $b$ in expected signed coefficient mass. The expected number of free variables after both restrictions is
\begin{equation}
m=\Theta_d\left(\frac{nb}{\ell^{d-2}}\right).
\label{eq:worst-heuristic}
\end{equation}
Moment bounds and point-list counting give expected maximum bounded-width CNF correlation $2^{-\Omega(m)}$. An exact circuit covers a constant fraction of the cube by at most $S^{O_d(1)}$ monochromatic CNFs, forcing $m=O_d(\ell)$. Taking $b=\Theta_d(\log n)$ and solving this relation gives $\ell=\Theta_d((n\log n)^{1/(d-1)})$.

The trilinear function and the evaluator in \cite{OpenAI26} provide moments through random parameter inputs. Inverse trace provides them through translations in a binary field, and Weil bounds for additive character sums. Its rational character sums remain controlled after conditioning on a coordinate subcube, paying explicitly for its codimension. This robustness is used in the average-case proof.

\paragraph{From exact computation to small correlation.}
For an exact circuit with OR output, the satisfying assignments of every CNF child lie in one level set of the target. Its density is therefore an absolute correlation. A circuit that makes errors has no such property: the accepting sets of its CNF children may overlap and contain inputs from both level sets of the target. Bounding each child alone does not bound the top OR's correlation.

We instead switch the entire restricted circuit down to depth two. For a circuit with bottom fan-in $O(\ell)$, $d-2$ switching stages followed by CNF width reduction leave an expected number of free variables
\begin{equation}
m=\Theta_d\left(\frac{n\log\ell}{\ell^{d-1}}\right),
\label{eq:average-heuristic}
\end{equation}
using $b=\Theta(\log\ell)$. The switching lemma gives failure probability only $2^{-\Omega(\ell)}$, which is insufficient when $m\gg\ell$. Common partial decision trees allow a much smaller failure probability $2^{-\Omega_d(m)}$.

The cost of a common partial decision tree is paid through an exact expansion. At a leaf, let $P$ indicate its path and let $H$ be a residual CNF. Then $PH$ is still a CNF of the same width: the path contributes unit clauses, irrespective of the path length. A DNF leaf is written as $P-PH$. Thus a tree of depth $R$ gives signed coefficient mass at most $2^{R+1}$ in bounded-width CNFs. Choosing $R$ as a sufficiently small depth-dependent fraction of $m$ leaves an exponentially decaying correlation.

Finally, simply replacing bottom gates of large fan-in by their typical values introduces an error of order $S2^{-k}$, again too large when $m\gg\ell$. We remove it by a signed cofactor recursion. Each exceptional assignment fixes many coordinates and incurs its full probability. The bounded-width CNF estimate holds uniformly on large coordinate subcubes, so the recursion can continue until its residual probability is $2^{-\Omega(n)}$.

\paragraph{Organization and dependencies.}
\Crefrange{sec:partition}{sec:inverse} prove the point-list theorem and the inverse-trace restriction estimate. \Cref{sec:average-proof} proves the correlation theorem, with detailed switching parameters and parameter checks in \Crefrange{app:switching}{app:parameters}. \Cref{sec:worst-proof} gives the common worst-case reduction; \Cref{app:other-targets} supplies the two other targets. \Cref{sec:upper} gives the complementary nonuniform upper bounds for inverse trace. \Cref{sec:limits} compares these bounds and discusses the limits of the lower-bound method. The external inputs are the switching lemma and multi-switching lemma of H\aa stad (the latter in Tal's formulation), rational additive-character bounds, and the trilinear restriction moments of Chukhin. Statements and locations are given where used. All other reductions are proved here.

\section{An exact decomposition by lists of satisfying assignments}
\label{sec:partition}
A clause is a disjunction of literals, and a CNF is a conjunction of clauses. Its width is the maximum number of variables in a clause. We remove tautological clauses and repeated literals. We identify a CNF with its Boolean indicator or its set of satisfying assignments, as appropriate. The constant-zero and constant-one functions are allowed, represented by an empty clause and an empty conjunction, respectively. The closure in Eqn.~\eqref{eq:closure-intro} forbids exactly the coordinate patterns of width at most $b$ absent from $A$.

\begin{theorem}[Quantitative point-list partition]
\label{thm:partition}
Let $m\ge1$, $b\ge1$ be integers and $C\ge2$ a real number. Put
\[
\eps_C=h_2(1/C)+1/C,\qquad
L_C(m,b)=\ceil{2^{Cb}\bigl[b\ln(2m+1)+1\bigr]},
\]
where $h_2(u)=-u\log u-(1-u)\log(1-u)$. Every width-$b$ CNF $H$ has an exact disjoint decomposition
\begin{equation}
\ind_H=\sum_{i=1}^R\ind_{\cl_b(A_i)},\qquad
R\le2^{\eps_Cm},\qquad 1\le|A_i|\le L_C(m,b).
\label{eq:partition}
\end{equation}
Each $A_i$ is a list of satisfying assignments of $H$. The empty set has the empty decomposition. For $m=0$ the assertion holds with one singleton list for the nonempty cube.
\end{theorem}

\begin{proof}
Fix a deterministic order of partial assignments on at most $b$ coordinates. At a nonempty current set $S$, query the first pattern $\rho$ with
\[
0<\Prb_{X\sim U_S}[X\text{ extends }\rho]<2^{-Cb},
\]
if one exists. Split $S$ into the true child $S\cap[\rho]$ and the false child $S\setminus[\rho]$. Both children are nonempty and smaller than $S$. The true child adds unit clauses; the false child adds the clause forbidding $\rho$. Every state remains a width-$b$ CNF on the original $m$ coordinates. The process terminates and its leaves partition $H$.

Every true branch decreases $\log|S|$ by more than $Cb$. Hence a nonempty terminal path has at most $m/(Cb)$ true branches. Let $\sigma$ be the union of their partial assignments. It is consistent, every terminal satisfying assignment extends it, and $|\operatorname{dom}\sigma|\le m/C$. Here $\operatorname{dom}\sigma$ means the domain of the partial assignment \(\sigma\).

Distinct leaves have distinct $\sigma$. At their first divergence, one takes the true branch of a pattern $\rho$. Its fingerprint $\sigma$ extends $\rho$. If the other leaf had the same fingerprint, all its satisfying assignments would extend $\rho$, contrary to its false branch. Thus one partial assignment records enough information to distinguish the leaves; neither query order nor negative answers need be stored. The leaf count is at most
\[
\sum_{j\le\floor{m/C}}2^j\binom mj
\le2^{m/C+mh_2(1/C)}.
\]
We used the binomial entropy estimate $\sum_{j\le r}\binom mj\le2^{mh_2(r/m)}$ for integer $0\le r\le m/2$ and monotonicity of $h_2$ on $[0,1/2]$.

At a terminal set $J$, every occurring pattern on at most $b$ coordinates has probability at least $\theta=2^{-Cb}$. There are at most $(2m+1)^b$ such patterns: encode the sorted literals and pad the remaining slots. With $L=L_C(m,b)$ independent uniform samples from $J$, the probability of missing any occurring pattern is at most
\[
(2m+1)^b e^{-L\theta}\le e^{-1}<1.
\]
Therefore we can choose a covering list $A$ with $|A| \leq L$. Every local pattern of a point of $J$ is in $A$, so $J\subseteq\cl_b(A)$. Conversely, a point outside $J$ violates one of its width-$b$ clauses. That falsifying pattern occurs in no satisfying assignment of $J$, hence in no row of $A$, and the point is outside $\cl_b(A)$. Thus $J=\cl_b(A)$ exactly.
\end{proof}

Since $\eps_C\to0$, the theorem implies \Cref{thm:partition-informal}. We will use
\begin{equation}
\eps_{512}<1/32,\qquad \eps_{2048}<1/128,
\label{eq:eps-constants}
\end{equation}
which follow from $h_2(u)\le u\log(e/u)$. Lists of lengths one through $L$ define fewer than $2^{mL+1}$ tests for $m\ge1$, regardless of their clause counts.

\paragraph{A small example.}
For $b=1$, the closure retains exactly the coordinate values present in the rows. The list $A=\{000,110\}$ has closure $\{000,010,100,110\}$: the first two bits are unrestricted and the last is zero. For $b=2$, the same list has closure $\{000,110\}$, since the first two bits must agree. The width parameter specifies which local patterns a list must preserve.

The decomposition also adapts to the number of satisfying assignments.
\begin{corollary}[Density refinement]
\label{cor:density}
If $H\ne\varnothing$ and $h=\log|H|$, the partition can be chosen with
\[
\log R\le B_C(h):=mh_2\left(\frac{h}{Cm}\right)+\frac hC.
\]
If every allowed point-list test $J$ satisfies $|\E[gJ]|\le\beta\sqrt{\mu_J}$, where $\mu_J=\E J$, then
\[
|\E[gH]|\le\beta\,2^{B_C(h)/2}\sqrt{\mu_H}.
\]
\end{corollary}
\begin{proof}
The logarithmic cardinality budget starts at $h$, so each fingerprint has at most $h/C$ fixed coordinates. Count as above. Disjointness and Cauchy--Schwarz give $\sum_i\sqrt{\mu_{J_i}}\le\sqrt{R\mu_H}$.
\end{proof}

\section{Restriction transfer without a clause-count loss}
\label{sec:transfer}
For a real function $G$ define
\[
\Mass_b(G)=\inf\left\{\sum_i|c_i|:\ G=\sum_i c_iJ_i,\quad J_i\text{ width-}b\text{ CNFs}\right\}.
\]
Equivalently, $\Mass_b(G)$ is the minimum $\ell_1$ norm of the coefficient vector over all exact expansions into width-$b$ CNFs. It is the quantity denoted $\operatorname{size}_b(G)$ by Chukhin~\cite[Section 2]{Chukhin26}. Then $|\E[gG]|\le\Mass_b(G)\kappa_b(g)$. Multiplication by a conjunction of unit clauses does not increase this mass. A $p$-random restriction independently keeps a variable free with probability $p$ and otherwise fixes it to a fair bit. Composition of independent $q$- and $p$-restrictions has the $qp$ product distribution.

\begin{lemma}[Expected CNF mass, Chukhin~\cite{Chukhin26}]
\label{lem:mass}
If $H$ is a width-$k$ CNF, $k\ge b\ge1$, and
\[
\vartheta=\sum_{j>b}\binom kj\left(\frac{2p}{1-p}\right)^j\le1,
\]
then $\E_\rho\Mass_b(H_\rho)\le1$. In particular this holds for $p=b/(64k)$.
\end{lemma}

\begin{proof}
We include the mass argument to make its independence of the clause count explicit. Order the clauses, and let $H_i$ be the conjunction of the first $i$. For the next clause $D$, call it wide if no literal is fixed true and more than $b$ variables remain free. Let $v_D$ be the conjunction falsifying the remaining literals, and let $\rho_D$ extend $\rho$ by these assignments. On a wide restriction,
\[
(H_i)_\rho=(H_{i-1})_\rho-v_D(H_{i-1})_{\rho_D}.
\]
The last cofactor is extended to the original cube of free variables by ignoring its additional fixed coordinates. The mass is at most the sum of the two previous masses. If $\rho$ falsifies $D$, the new mass is zero. In every other nonwide case, adding $D_\rho$ to each summand does not increase mass.

Each preimage of a restriction $\sigma$ falsifying $D$ is obtained by freeing $j>b$ of its at most $k$ clause variables. Its product probability is $(2p/(1-p))^j$ times that of $\sigma$. Summing the extra wide-case mass therefore costs at most $\vartheta$ times the previous mass on restrictions falsifying $D$, precisely the mass removed in that case. Expected mass cannot increase as clauses are added. It starts at one for $H_0=1$.

For $p=b/(64k)$, $2p/(1-p)\le4p$, and each summand with $j>b$ is at most $(eb/(16j))^j\le4^{-j}$. Their sum is below one.
\end{proof}

\begin{lemma}[A simultaneous good event]
\label{lem:transfer}
Let $k\ge b\ge1$ be integers and let $\gamma,\Delta\ge0$. Let $\phi$ have any distribution on outer restrictions, and conditional on $\phi$ let $\rho$ be a fresh $p$-restriction with $p=b/(64k)$. If $|g|\le1$ and $\kappa_b(g_{\phi\rho})\le\gamma$ outside a joint event of probability at most $\Delta$, then
\[
\E_\phi\kappa_k(g_\phi)\le\gamma+\Delta.
\]
\end{lemma}
\begin{proof}
Fix $\phi$ and a width-$k$ CNF $H$ before drawing $\rho$. Averaging a restriction and filling its free variables reproduces the uniform input distribution. Thus
\begin{align*}
|\E[g_\phi H]|&\le\E_\rho|\E[g_{\phi\rho}H_\rho]|\\
&\le\gamma\E_\rho\Mass_b(H_\rho)+\Prb_\rho[\mathrm{bad}\mid\phi]\\
&\le\gamma+\Prb_\rho[\mathrm{bad}\mid\phi].
\end{align*}
On a bad restriction we use $|\E[g_{\phi\rho}H_\rho]|\le1$. The right side is independent of $H$, so we can maximize over $H$ and then average $\phi$.
\end{proof}

The order matters: the resulting CNF may depend on all outer circuit restrictions. It is selected before the fresh inner restriction used in \Cref{lem:transfer}. We never assume independence between the good event and coefficient mass.

\section{Inverse trace against every restricted CNF}
\label{sec:inverse}
Let $E=\F_{2^N}$ with a fixed binary basis, $\psi(x)=(-1)^{\Tr(x)}$, and $f(x)=\psi(c\inv(x))$ for $c\ne0$. Inverse trace corresponds to $c=1$. A coordinate subcube $V=x_0+W$ fixes some basis coordinates; its codimension $b_0$ is the number fixed. Expectations on subcubes are normalized.

\begin{lemma}[Additive-character sums for products of shifts]
\label{lem:shifts}
For distinct $u_1,\ldots,u_h\in E$, $h\ge1$, and every $a\in E$,
\begin{equation}
\abs{\E_{x\in E}\psi(ax)\prod_{i=1}^h f(x+u_i)}\le8h\,2^{-N/2}.
\label{eq:complete-shifts}
\end{equation}
If $u_i\in W$ and $V=x_0+W$ has codimension $b_0$, then
\begin{equation}
\abs{\E_{x\in V}\prod_{i=1}^h f(x+u_i)}\le8h\,2^{b_0-N/2}.
\label{eq:face-shifts}
\end{equation}
\end{lemma}
\begin{proof}
Off its poles the summand in Eqn.~\eqref{eq:complete-shifts} is $\psi(r(x))$ for
\[
r(X)=aX+\sum_{i=1}^h\frac{c}{X+u_i}.
\]
Its finite poles are simple with nonzero residues. It cannot be $R^2-R$ over the algebraic closure, since every pole of such a nonconstant expression has even order. Apply the rational additive-character estimate, for example Fu--Wan~\cite[Theorem 4.6, version 3]{FuWan}, with extension degree one, trivial multiplicative character on $\prod_i(X+u_i)$, and additive argument $r$. In their notation $D_1=D_3=h$, $D_4=0$, and $D_2\le1$, giving an unnormalized bound at most $2h\,2^{N/2}$ off the poles. Restoring the at most $h$ values at the poles under our convention $\inv(0)=0$ proves Eqn.~\eqref{eq:complete-shifts} with slack.

Let $W^\perp$ be the annihilator of $W$ under the nondegenerate trace pairing. Orthogonality gives
\[
\ind_V(x)=2^{-b_0}\sum_{a\in W^\perp}\psi(a(x-x_0)).
\]
Normalized conditioning is therefore a sum of $2^{b_0}$ twisted full-field expectations. Apply Eqn.~\eqref{eq:complete-shifts} to each.
\end{proof}

\begin{lemma}[Even moments under random translation]
\label{lem:moment}
Let $L_0:\F_2^v\to W$ be injective and let $J$ be a Boolean indicator of density $\mu$. For every integer $t\ge1$,
\begin{equation}
\E_{A\sim U_V}\abs{\E_z f(A+L_0z)J(z)}^{2t}
\le(2t-1)!!(\mu2^{-v})^t+16t\,2^{b_0-N/2}\mu^{2t}.
\label{eq:moment}
\end{equation}
\end{lemma}
\begin{proof}
Expand over $2t$ independent uniform points $z$. Tuples in which each point has even multiplicity are covered by the $(2t-1)!!$ pairings of positions. Each pairing has total $J$-weight $(\mu2^{-v})^t$. Every other tuple leaves a nonempty set of at most $2t$ distinct shifts of odd multiplicity. Apply Eqn.~\eqref{eq:face-shifts}; all tuple weights are nonnegative and their total is $\mu^{2t}$.
\end{proof}

The paired term behaves like a moment bound for limited independence. The second term is the character-sum error term. Our list count must fit within the exponential bound supplied by the latter.

\begin{proposition}[Expected maximum CNF correlation on large subcubes]
\label{prop:inverse}
Fix a coordinate subcube $V$ of codimension $b_0\le N/4$ and dimension $v_0=N-b_0$. Let $0<q\le1$ be deterministic and let $k\ge b\ge2$ be integers. Set
\begin{equation}
p=\frac b{64k},\quad m=v_0qp,\quad
L=\ceil{2^{512b}\bigl[b\ln(4m+1)+1\bigr]}.
\label{eq:inverse-parameters}
\end{equation}
If
\begin{equation}
m\ge2b,\qquad m\ge16\log(8L),\qquad mL\le N/64,
\label{eq:inverse-hypotheses}
\end{equation}
then for a $q$-random restriction $\phi$ of $V$,
\begin{align}
\E_\phi\kappa_k(f_{V,\phi})&\le2^{-3m/64}+\Delta,\label{eq:inverse-result}\\
\Delta&\le2\cdot2^{-m/8}+2^{1-3mL/4}+128L\,2^{-3N/16}.\nonumber
\end{align}
The maximum over every width-$k$ CNF is taken before averaging $\phi$.
\end{proposition}

\begin{proof}
Draw a fresh inner $p$-restriction $\rho$. The number $v$ of free variables under $\sigma=\phi\rho$ has expectation $m$. Chernoff bounds give $v\in[m/2,2m]$ except with probability at most $2e^{-m/8}\le2\cdot2^{-m/8}$.

Fix a restriction whose number of free variables lies in this interval. First translate the target by a uniform element of $W$; its offset becomes uniform in $V$. In \Cref{lem:moment} take $t=4L$ and threshold $\beta=2^{-v/8}$. Union over fewer than $2^{vL+1}$ list tests and use $(2t-1)!!\le(2t)^t$. Their simultaneous failure probability is at most
\begin{equation}
2^{1-2vL+4L\log(8L)}+
128L\,2^{b_0-N/2+2vL}
\le2^{1-3vL/2}+128L\,2^{b_0-N/2+2vL},
\label{eq:list-union}
\end{equation}
because $v\ge8\log(8L)$. The list-length bound $L$ dominates $L_{512}(v,b)$. On success, \Cref{thm:partition} and Eqn.~\eqref{eq:eps-constants} imply
\[
\kappa_b(f\text{ on this subcube})\le2^{v/32}\beta
=2^{-3v/32}\le2^{-3m/64}.
\]
The second term of Eqn.~\eqref{eq:list-union} is at most $128L2^{-3N/16}$, since $b_0\le N/4$ and $2vL\le4mL\le N/16$. The first is at most $2^{1-3mL/4}$.

We now remove the translation average. Translation within $W$ flips the unfixed binary coordinates of $V$. It preserves the product distribution of random restrictions and maps the class of all width-$b$ CNFs on the surviving bits to itself. Thus the bad-event probability just bounded is the same for each fixed translate of the target. Its average bound holds for the untranslated target as well. This does not say that every small subcube is good. It gives a joint good event for $(\phi,\rho)$ under their unconditional product distribution. Apply \Cref{lem:transfer}.
\end{proof}

The case $V=E$ will suffice for worst-case hardness. The uniformity over initial subcubes of codimension up to $N/4$ is needed to remove the restriction on bottom fan-in in the correlation argument.

\section{Controlling the whole circuit}
\label{sec:average-proof}
We first work at a tower length $N$ with $f=(-1)^{I_N}$. Since inversion permutes $E$ and trace is a nonzero linear functional, $\E f=0$. Consequently, for Boolean $C$,
\begin{equation}
\Corr(I_N,C)=2|\E[fC]|.
\label{eq:boolean-sign}
\end{equation}
We prove bounds on the right, including bounds on subcubes where $f$ need not be balanced.

\subsection{Guarded CNFs and common trees}
\begin{lemma}[Guarding leaves preserves width]
\label{lem:guards}
If a Boolean function $G$ has a restriction tree of depth at most $R$, with each residual leaf a width-$k$ CNF or DNF, then
\[
\Mass_k(G)\le2^{R+1},\qquad
|\E[gG]|\le2^{R+1}\kappa_k(g).
\]
\end{lemma}
\begin{proof}
Let $P_\lambda$ be the conjunction of literals specifying the root-to-leaf path. We call the product $P_\lambda H_\lambda$ a \emph{guarded CNF}. A CNF leaf contributes $P_\lambda H_\lambda$. For a DNF leaf write its value as $1-H_\lambda$ with $H_\lambda$ a CNF, and contribute $P_\lambda-P_\lambda H_\lambda$. Extend residual functions by ignoring path coordinates. All atoms are width-$k$ CNFs, since a path is a conjunction of units. Paths partition the cube and each of at most $2^R$ leaves contributes at most two atoms.
\end{proof}

We use the following switching inputs. A width-$k$ CNF or DNF under a $p$-restriction has a decision tree of depth at most $t$ except with probability $(5pk)^t$~\cite{Hastad89}. It then has both width-$t$ forms with at most $2^t$ terms or clauses. H\aa stad's multi-switching lemma~\cite{Hastad14}, in the formulation of Tal~\cite[Definition 14 and Lemma 15]{Tal17}, states that a family of $M$ depth-two circuits of bottom fan-in at most $k$ has, except with probability $M(24pk)^D$, a common $\ceil{\log(2M)}$-partial decision tree of depth at most $D$. At each leaf, each member has an ordinary decision tree of the stated residual depth. We use the constant $24$ from the cited formulation.

Here and below, circuits are initially alternating and layered. Constants and unary gates are retained, and original layer labels are retained under cofactors. These conventions keep the circuit class closed under the operations in the proof. \Cref{app:uniform} normalizes general circuits at polynomial size cost.

\begin{proposition}[Two switching estimates]
\label{prop:switching}
Fix $d\ge3$, $S\ge N$, $\ell=\log S$, and $s=d-2$. For layered alternating circuits of depth at most $d$, size at most $S$, and bottom fan-in at most
\[
k_0=\ceil{4\log(2S)},
\]
there are deterministic parameters $k=\Theta_d(\ell)$ and $q=\Theta_d(\ell^{-s})$ with the following properties. For every initial subcube $V$ and $|g|\le1$,
\begin{equation}
|\E_V[gC]|\le s2^{-3\ell}+2\E_\phi\kappa_k(g_{V,\phi}).
\label{eq:ordinary-switch}
\end{equation}
Also, for any positive integer $D_0$, with $R_s=(2^s-1)D_0$ and $A=6^d$,
\begin{equation}
|\E_V[gC]|\le s2^{A\ell-2D_0}
+2^{R_s+1}\E_\phi\kappa_k(g_{V,\phi}).
\label{eq:common-switch}
\end{equation}
Here $\phi$ has the $q$-random restriction distribution. It is drawn even after simplification failures; neither $q$ nor $k$ depends on a successful path.
\end{proposition}

The full construction is in \Cref{app:switching}. Its key feature is that $s$ switching stages use fixed survival probabilities $\Theta_d(1/\ell)$. With common trees the depth bound at stage $i$ pays for all previous leaves, yet the final tree depth stays at most $(2^s-1)D_0$. \Cref{lem:guards} then gives Eqn.~\eqref{eq:common-switch}. At no point do we assert a product distribution of random restrictions after conditioning on a successful leaf.

\subsection{Circuits with bounded bottom fan-in}
Define
\[
\mathcal R=\frac{N\log\ell}{\ell^{d-1}}.
\]
Take $b=\floor{\log k/2048}$ and $p=b/(64k)$. On any initial subcube of codimension less than $N/4$, the expected number of free variables in \Cref{prop:inverse} is
\[
m=(\dim V)qp=\Theta_d(\mathcal R).
\]
For a sufficiently small $c_*>0$, uniformly in
\begin{equation}
\log N\le\ell\le c_*N^{1/(d-1)},
\label{eq:average-range}
\end{equation}
the hypotheses in Eqn.~\eqref{eq:inverse-hypotheses} hold and
\begin{equation}
\E_\phi\kappa_k(f_{V,\phi})\le5\cdot2^{-3m/64}.
\label{eq:inverse-simplified}
\end{equation}
Indeed the list-length bound is $L=O_d(\ell^{1/4}\log\ell\log N)$, $m$ is at least a sufficiently large multiple of $\log\ell$, and $mL=o(N)$ even at $\ell=\log N$. \Cref{app:parameters} checks this uniformly, including both size endpoints.

When $m$ is at most a sufficiently large depth-dependent multiple of $\ell$, use Eqn.~\eqref{eq:ordinary-switch}. Both its terms are $2^{-\Omega_d(m)}$. For larger $m$, choose $D_0=\floor{m/J}$ with $J$ a sufficiently large depth constant. Then $R_s\le m/4096$ and the failure term in Eqn.~\eqref{eq:common-switch} is also $2^{-\Omega_d(m)}$. Combining the cases proves:

\begin{proposition}[Correlation for circuits with bounded bottom fan-in]
\label{prop:narrow}
There are constants $c_*,\theta_d>0$ such that, throughout the range in Eqn.~\eqref{eq:average-range}, every circuit in \Cref{prop:switching} satisfies
\[
|\E_V[fC]|\le\Gamma:=2^{-\theta_d\mathcal R}
\]
on every coordinate subcube $V$ of codimension less than $N/4$.
\end{proposition}

\subsection{Removing the restriction on bottom fan-in}
The following lemma replaces an additive truncation error by recursion through cofactors. It is useful precisely because \Cref{prop:narrow} holds on every large initial subcube.

\begin{lemma}[Signed cofactor transfer]
\label{lem:cofactor}
Let $|g|\le1$ on $N$ coordinates. Suppose all layered circuits of depth at most $d$, size at most $S$, and bottom fan-in at most $k$ satisfy $|\E_V[gC]|\le\Gamma$ on every coordinate subcube $V$ of codimension less than an integer $B\le N$. If
\[
a_0=2S2^{-k}<1,\qquad 2S2^{-k/2}\le1,
\]
then the same circuits with unrestricted bottom fan-in satisfy
\[
|\E[gC]|\le\frac{\Gamma}{1-a_0}+2^{-B/2}.
\]
\end{lemma}
\begin{proof}
Let $B_b$ be the maximum of $|\E_V[gC]|$ over all subcubes $V$ with $b$ fixed coordinates and all circuits $C$ in the class with unrestricted bottom fan-in. On a subcube with $b<B$, list its bottom gates whose fan-ins $w_i$ exceed $k$, removing repetitions and simplifying contradictory literals first. Replace each OR by one and each AND by zero, in sequence. The final circuit has bottom fan-in at most $k$. Gate $i$ differs from its replacement only on the coordinate subcube on which all its literals are false (for an OR gate) or all are true (for an AND gate), of relative measure $2^{-w_i}$.

On that subcube the circuit difference is a difference of two Boolean cofactor circuits, each of depth at most $d$ and size at most $S$. Sharing the replaced gate among several parents does not duplicate either cofactor circuit. Telescoping therefore gives
\begin{equation}
B_b\le\Gamma+\sup_C 2\sum_i2^{-w_i}B_{b+w_i}.
\label{eq:cofactor-recursion}
\end{equation}
There is no first-exception prefix condition in this identity.

Descending induction on $b$ gives
\[
B_b\le\frac{\Gamma}{1-a_0}+2^{-(B-b)/2}.
\]
For $b\ge B$ the right side is at least one. For $b<B$, the constant contribution from Eqn.~\eqref{eq:cofactor-recursion} is at most $\Gamma+a_0\Gamma/(1-a_0)$. The remaining contribution is at most
\[
2^{-(B-b)/2}\,2\sum_i2^{-w_i/2}\le2^{-(B-b)/2}.
\]
Each recursive call fixes fresh coordinates, so the induction is well founded. Set $b=0$.
\end{proof}

Apply the lemma with $k=k_0$ and $B=\floor{N/4}$. Then $a_0<1/2$ and the second hypothesis holds. \Cref{prop:narrow} yields
\[
|\E[fC]|\le2\cdot2^{-\theta_d\mathcal R}+2^{-\floor{N/4}/2}
\le2^{-\theta'_d\mathcal R}.
\]
Here $\mathcal R=o(N)$ uniformly over the range in Eqn.~\eqref{eq:average-range}, and its minimum in that range tends to infinity. This proves the correlation estimate for layered circuits at tower lengths. Circuit normalization costs $O_d(S_0^2)$ gates, which we absorb using budget $S=S_0^3$. Eqn.~\eqref{eq:boolean-sign} and padding from $N$ to $n<2N$ change only depth-dependent constants. The uniform field construction, normalization, and padding details are in \Cref{app:uniform}; they complete \Cref{thm:average}.

\section{Worst-case hardness through monochromatic covers}
\label{sec:worst-proof}
For exact computation we can stop depth reduction one layer earlier. We use H\aa stad's depth reduction in the form stated by Chukhin~\cite[Lemma 2.3]{Chukhin26}: for fixed $d\ge3$ and $S\ge M\ge2$, there are common parameters
\[
k=\Theta_d(\log S),\qquad q=\Theta_d((\log S)^{-(d-3)})
\]
such that a $q$-restriction of a size-$S$, depth-at-most-$d$ circuit with OR output is an OR of at most $S^{A_d}$ width-$k$ CNFs with probability at least $3/4$. The same parameters work for all such circuits on $M$ bits.

A CNF is \emph{monochromatic} for a target if the target is constant on its set of satisfying assignments.

\begin{lemma}[Covering by monochromatic CNFs]
\label{lem:cover}
Let $G$ be Boolean, $g=(-1)^G$, and suppose both level sets of $G$ have density at least $\delta$. If a size-$S$, depth-at-most-$d$ circuit computes $G$, then for the parameters above,
\[
\delta\le\frac14+S^{A_d}\E_\phi\kappa_k(g_\phi).
\]
\end{lemma}
\begin{proof}
Complement all gates and literals if needed to give an OR output computing $G$ or $1-G$. On a successful depth reduction, the satisfying assignments of each top CNF lie in the corresponding level set of $G$. Its density equals its absolute correlation with $g_\phi$. Sum these densities over the at most $S^{A_d}$ children. On failure use one, then average $\phi$. Restriction averaging preserves the unconditional output density.
\end{proof}

\paragraph{Inverse trace.}
At a tower length $N$, put $\ell=c(N\log N)^{1/(d-1)}$, $S=2^\ell$, and
\[
b=\floor{\frac{\log N}{8192(d-1)}},\quad
p=\frac b{64k},\quad m=Nqp.
\]
Then
\[
m=\Theta_d\bigl(c^{-(d-2)}(N\log N)^{1/(d-1)}\bigr),\quad
L\le N^{1/[16(d-1)]}O_d((\log N)^2).
\]
Thus $mL=o(N)$, $m\gg\log L$, and $m\gg b$. \Cref{prop:inverse} with $V=E$ gives
\[
\E_\phi\kappa_k(f_\phi)\le5\cdot2^{-3m/64}
\]
eventually. Since $I_N$ is balanced, \Cref{lem:cover} would imply
\[
\frac12\le\frac14+5S^{A_d}2^{-3m/64}.
\]
But $m/\ell=\Theta_d(c^{-(d-1)})$, so a sufficiently small $c$ makes the last term tend to zero. Padding proves the claim for $F_n$.

\paragraph{The other two targets.}
For the trilinear function, choose one coefficient slice using Chukhin's restriction moments and union over our point-list tests. For the evaluator, fix an irreducible modulus for the analysis, use the original linear hash, and obtain $t$-wise independent signs by polynomial interpolation, where $t$ is the number of polynomial coefficients. In both cases the slice is selected before the circuit is chosen, and its two level sets each have density at least $1/3$. The common point-list transfer gives $\E_\phi\kappa_k(g_\phi)\le5\cdot2^{-\Omega(m)}$ at the scale in Eqn.~\eqref{eq:worst-heuristic}. \Cref{lem:cover} finishes with $\delta=1/3$. \Cref{app:other-targets} gives the full moments, list counts, and input accounting. This completes \Cref{thm:worst}.

\section{Explicit upper bounds for inverse trace}
\label{sec:upper}
We prove \Cref{thm:upper} by evaluating a product of Frobenius maps over an auxiliary field of polynomial size. Evaluation and interpolation over smaller fields are standard tools in constant-depth finite-field arithmetic~\cite{HV05}. The calculation below implements the arithmetic directly at the required Boolean depth, rather than first constructing a logarithmic-depth circuit and then compressing it. The qualitative existence of subexponential-size circuits at some fixed depth also follows from~\cite{HV05} for its explicit field representation; the depth allocations below are supplied here.

All circuits in this section are nonuniform. Field presentations, changes of basis, evaluation points, and lookup tables are fixed circuit data. The final circuits contain only unbounded-fan-in AND/OR gates and input negations. We count both gates and wires.

\subsection{Enumerating intermediate states}
\begin{lemma}[Computations with a bounded number of state bits]
\label{lem:upper-state}
Consider a deterministic computation that reads $T$ consecutive input blocks of $b$ bits each and maintains a $w$-bit state, where $T,b,w\ge1$. Its initial state is fixed, each transition is an arbitrary fixed function of the state and the current block, and its output is an arbitrary fixed predicate of the final state. Put $L=Tb$. For every fixed integer $k\ge1$, the output has a depth-at-most-$(k+1)$ AND/OR circuit of size
\begin{equation}
2^{O_k\left(L^{1/k}w^{1-1/k}+b+w+\log(T+2)\right)}.
\label{eq:upper-state-size}
\end{equation}
Both an AND-top and an OR-top circuit are available, for the output and its complement. The same bound holds for the number of wires.
\end{lemma}
\begin{proof}
For $k=1$, use a full CNF or DNF on the $L$ input bits. Suppose $k\ge2$ and $1\le w<L$. Set $g=\ceil{(L/w)^{1/k}}$, and first assume $g\le T$.

Split the block sequence into at most $g$ consecutive intervals of nearly equal length, and repeat this splitting for $k-1$ levels. To test a segment with a prescribed initial state and a prescribed final-state predicate, enumerate the states at its interval boundaries. For each boundary-state tuple, conjoin the corresponding interval tests; then take the OR over tuples. There are at most $2^{gw}$ tuples at a node. At a leaf, the segment has at most $\ceil{T/g^{k-1}}$ blocks, so its test has a full CNF on at most $L/g^{k-1}+b$ input bits.

Each interval test is again a computation of the same kind. Complementing its final-state predicate complements the test, so De Morgan's law gives an AND-top implementation whenever an OR-top implementation is available, and conversely. At each internal node, choose AND-top circuits for the interval tests and merge their top gates with the conjunction for a boundary-state tuple. Each recursion level therefore adds one layer. Starting from depth two at the leaves gives total depth $k+1$. Complementing the final predicate and applying De Morgan also gives both root types for both output polarities.

The number of levels is fixed. Counting copies, boundary tuples, and leaf truth tables bounds the logarithm of the size by
\[
O_k\left(gw+\log(g+1)+\frac{L}{g^{k-1}}+b+\log(T+2)\right).
\]
Here $gw=O_k(L^{1/k}w^{1-1/k}+w)$ and $L/g^{k-1}\le L^{1/k}w^{1-1/k}$. Leaf literal wires add only their input-length factor, and internal conjunctions add at most $g$ wires per tuple, so the same estimate covers wires.

For completeness, if $w\ge L$ or $T=1$, the full truth table suffices. If $g>T$, then $(L/w)^{1/k}>T$, which gives $Tw<b$ for $k\ge2$. A single split into individual blocks has logarithmic size $O(Tw+b+\log(T+2))$ and depth three, within the claimed bounds. This covers the rounding cases.
\end{proof}

\subsection{Representing inverse trace by evaluations}
Let $E=\F_{2^N}$, with $N\ge2$, in any fixed binary basis. Fix a polynomial presentation $E\cong\F_2[Z]/(p)$, where $p$ is irreducible of degree $N$. For $1\le j\le N-1$, let $f_j(Z)$ be the degree-below-$N$ representative of $x^{2^j}$. Its coefficients are binary linear functions of the original input bits. Define
\[
P_x(Z)=\prod_{j=1}^{N-1}f_j(Z).
\]
This polynomial has degree at most $D=(N-1)^2$, and
\[
P_x\bmod p=\inv(x),
\]
because $\sum_{j=1}^{N-1}2^j=2^N-2$. The identity also holds at zero. Reduction modulo $p$ followed by trace is a fixed binary linear functional of the coefficients of $P_x$; denote it by $\Lambda$.

Put $M=D+1$ and choose an auxiliary field $K=\F_{2^s}$ with $s=\ceil{\log M}$. Choose $M$ distinct points $\alpha_1,\ldots,\alpha_M\in K$. The auxiliary field need not embed in $E$. Interpolation determines $P_x$ from the values
\[
v_\ell(x)=P_x(\alpha_\ell)=\prod_{j=1}^{N-1}f_j(\alpha_\ell).
\]
To extend $\Lambda$ to the coefficients obtained by interpolation over $K$, choose a binary linear map $\pi:K\to\F_2$ with $\pi(1)=1$ and apply it to each coefficient before applying $\Lambda$. This agrees with $\Lambda$ on the actual polynomial, whose coefficients lie in $\F_2$. The resulting map from $K^M$ to $\F_2$ is binary linear, so there are fixed binary linear functionals $\lambda_\ell:K\to\F_2$ such that
\begin{equation}
I_N(x)=\bigoplus_{\ell=1}^{M}\lambda_\ell(v_\ell(x)).
\label{eq:upper-evaluation}
\end{equation}
In particular, the final parity has $M=O(N^2)$ scalar inputs.

\subsection{Three stages at adjustable depths}
Fix positive integers $h_1,h_2,h_3$ and a generator $\gamma$ of $K^\times$. For $y\ne0$, represent $\log_\gamma y$ by $s$ bits; assign logarithm zero to $y=0$ and keep a separate zero flag. All logarithm and exponentiation tables are on $O(\log N)$ bits, so their descriptions have polynomial size.

\paragraph{Evaluation and logarithm tables.}
For every $j,\ell$, the map $x\mapsto f_j(\alpha_\ell)$ is a fixed binary linear map into $K$. A computation reading the $N$ raw input bits can accumulate its value in an $s$-bit state. Any chosen bit of its logarithm, or its zero flag, is a predicate of the final state. Applying \Cref{lem:upper-state} with $L=T=N$ and $b=1$ gives depth $h_1+1$ and logarithmic size
\begin{equation}
O_{h_1}\left(N^{1/h_1}(\log N)^{1-1/h_1}+\log N\right).
\label{eq:upper-stage-a}
\end{equation}
The polynomial number of required output bits adds only $O(\log N)$ to the logarithmic size. This stage includes all Frobenius maps and basis changes.

\paragraph{Multiplication at each evaluation point.}
For each $\ell$, read the $N-1$ logarithms and zero flags as blocks of $s+1$ bits. Maintain their logarithm sum modulo $|K|-1$ and the OR of the zero flags, using $s+1$ state bits. If a zero flag has occurred, output zero; otherwise the terminal predicate outputs $\lambda_\ell(\gamma^a)$, where $a$ is the accumulated sum. The transitions may be defined arbitrarily on invalid logarithm encodings, which do not arise from the first stage. Here $L=(N-1)(s+1)$ and $b=w=s+1$. Thus \Cref{lem:upper-state} gives depth $h_2+1$ and logarithmic size
\begin{equation}
O_{h_2}\left(N^{1/h_2}\log N+\log N\right).
\label{eq:upper-stage-b}
\end{equation}
There are only $M$ parallel computations. Although their formal inputs are correlated outputs of the first stage, the circuit is exact on every assignment to those formal inputs, so substitution is valid.

\paragraph{The final parity.}
Apply \Cref{lem:upper-state} to parity on the $M$ scalar outputs in Eqn.~\eqref{eq:upper-evaluation}, using one state bit. This gives depth $h_3+1$ and logarithmic size
\begin{equation}
O_{h_3}\left(N^{2/h_3}+\log N\right).
\label{eq:upper-stage-c}
\end{equation}

\begin{proof}[Proof of \Cref{thm:upper}]
Each stage has alternating gate layers and is available with either root type, for both output polarities. At a stage boundary, implement each input literal of the next stage by a circuit whose root type agrees with that next stage's bottom gate. Merge the adjacent gates of the same type. Both polarities are available, so this introduces no internal negation. The two boundaries save one layer each, giving depth
\[
(h_1+1)+(h_2+1)+(h_3+1)-2=h_1+h_2+h_3+1.
\]
Combining Eqns.~\eqref{eq:upper-stage-a}--\eqref{eq:upper-stage-c} gives logarithmic size $O_{\mathbf h}(\Phi_{\mathbf h}(N))$. Gate sharing is allowed. Even if merging increases the wire count, the total number of distinct wires is at most quadratic in the number of gates and inputs before merging, which preserves this estimate.

The full DNF gives size $2^{O(N)}$ and depth two. Taking the best construction among the finitely many triples allowed at a fixed depth $d$ proves Eqn.~\eqref{eq:main-upper}. The allocations $(h_1,h_2,h_3)=(2,2,3)$ and $(2,2,4)$ give Eqn.~\eqref{eq:upper-examples}; choosing $(h,h,2h)$ gives Eqn.~\eqref{eq:upper-symmetric}. The case $N=1$ is immediate. Finally, $F_n$ ignores all but its first $N$ bits, where $N\le n<2N$, so the same circuits give the asserted bounds in $n$.
\end{proof}

Thus inverse trace has exact circuits of size $2^{o(n)}$ at each fixed depth $d\ge8$. For every fixed $\eps>0$, choosing an integer $h>1/\eps$ in Eqn.~\eqref{eq:upper-symmetric} gives size $2^{O_\eps(n^\eps)}$ at depth $4h+1$. These estimates give sufficient depths and sizes; they do not establish optimality or exclude exponential hardness at depths three through seven.

\section{Consequences and limits of the representation}
\label{sec:limits}
The point-list theorem concerns descriptions of sets. It allows an exponential number of pieces, can require exact counts of satisfying assignments to find them, and permits many clauses per piece. Its use here is to make simultaneous probabilistic estimates possible. \Cref{app:refinements} derives a bound on the correlation of inverse trace with width-$k$ CNFs.

There is an intrinsic list-length cost. If $J=\cl_b(A)\ne\varnothing$, $1\le b\le m$, and $|A|\le L$, then every $b$-coordinate projection of $J$ has at most $L$ values. Shearer's entropy inequality gives
\[
|J|\le L^{m/b}.
\]
Consequently any exact signed representation $\ind_H=\sum_i c_i\ind_{\cl_b(A_i)}$ with $|A_i|\le L$ satisfies
\begin{equation}
\sum_i|c_i|\ge2^{h-(m/b)\log L},\qquad h=\log|H|.
\label{eq:list-lower}
\end{equation}
\Cref{app:refinements} proves both statements. For high-entropy $H$, coefficient mass $2^{\eps m}$ forces $\log L\ge b(h/m-\eps)$. Exponential dependence on width is therefore unavoidable for this dictionary.

In the inverse-trace union bound, the character-sum error term requires $mL=O(N)$. Together with Eqn.~\eqref{eq:list-lower}, this limits the useful width to logarithmic size. The worst-case proof also needs $m=\Omega_d(\log S)$. At depth three these constraints yield $\log S$ of order $\sqrt{N\log N}$; changing a constant in the partition cannot turn this argument into an exponential lower bound.

There is a separate higher-depth limitation of this covering route. If the outer restriction keeps each bit with probability $q$, then on any cube of free variables of dimension $v$ a singleton is a width-one CNF of absolute correlation $2^{-v}$ with any sign function. Hence
\[
\E_\phi\kappa_k(g_\phi)\ge\E_\phi2^{-v}
=(1-q/2)^N=2^{-O(Nq)}.
\]
To make this smaller than $S^{-A_d}$ with $q=\Theta_d((\log S)^{-(d-3)})$ requires $(\log S)^{d-2}=O_d(N)$. This is a restriction on the method, not a general upper bound on circuit complexity. For the trilinear function a specific upper bound $2^{O_d(n^{2/d})}$ is known~\cite[Theorem 5.1]{Chukhin26}, precluding truly exponential lower bounds at fixed $d\ge3$ for that target. For inverse trace, \Cref{thm:upper} rules out a size lower bound $2^{\Omega(n)}$ at every fixed depth $d\ge8$. It leaves open the corresponding question at depths three through seven, and its upper bounds remain above the lower-bound scale of \Cref{thm:worst}. Distinguishing these upper bounds on the functions from limitations of the proof method is essential when assessing the remaining gap in Eqn.~\eqref{eq:inverse-sandwich}.

\section*{Acknowledgments}
This work was carried out during the authors' visit to the Simons
Institute for the Theory of Computing. The authors thank the Institute
for its hospitality and support during the program Pseudorandomness
and High-Dimensional Expansion in Fall 2026.

\paragraph{AI disclosure.}
The authors initiated this project by investigating properties of
the inverse-trace function relevant to circuit lower bounds.
OpenAI Codex was used substantially, through iterative exchanges
with the authors, to explore proof ideas and assist with exposition,
proof auditing, and reference checking. %

\clearpage
\phantomsection
\clearpage

\appendix
\section{Deterministic switching parameters}
\label{app:switching}
We prove \Cref{prop:switching}, including its handling of shared gates and shallow leaves. Put $s=d-2$, $U_0=S$, and recursively set
\begin{equation}
t_i=\ceil{\log(2U_{i-1})},\quad
k_i=\max\{k_{i-1},t_i\},\quad
p_i=\frac1{96k_{i-1}},\quad
U_i=2^{t_i+2}U_{i-1},\qquad 1\le i\le s.
\label{eq:budgets}
\end{equation}
These quantities depend only on $S,d$. Since $2^{t_i}\le4U_{i-1}$,
\begin{equation}
U_i\le16U_{i-1}^2\le U_{i-1}^6,\qquad U_i\le S^{6^i}.
\label{eq:size-budget}
\end{equation}
In particular $U_i\le2^{A\ell}$ for $A=6^d$.

At an active leaf of depth at least three, consider each second-layer subcircuit. Switching it to the opposite depth-two form permits its outer gate to merge with its parent, decreasing the whole circuit depth by one. An ordinary decision tree of depth at most $t_i$ supplies both forms with at most $2^{t_i}$ terms or clauses. Replacing all at most $U_{i-1}$ second-layer gates fits the size bound $U_i$. If a gate has several parents, all use the same replacement; only wires are duplicated. Each individual residual circuit therefore stays within $U_i$ gates.

A leaf already at depth at most two is frozen. It retains its current CNF or DNF under later random restrictions, without new tree queries. The maximum in the definition of $k_i$ ensures that this width remains allowed, including when $k_0>t_1$.

At every stage draw one fresh global $p_i$-restriction on all currently free coordinates, including variables absent from the simplified circuit. Continue drawing after any failure. The unconditional composition $\phi$ has restriction parameter
\begin{equation}
q=\prod_{i=1}^s p_i,\qquad k=k_s.
\label{eq:global-survival}
\end{equation}
For fixed $d$, both $k=\Theta_d(\ell)$ and $q=\Theta_d(\ell^{-s})$ follow directly from these parameters.

\subsection{The switching lemma}
Condition on any successful prior history. At stage $i$, union over at most $U_{i-1}$ active second-layer functions. Its conditional failure probability is at most
\[
U_{i-1}(5/96)^{t_i}
\le U_{i-1}16^{-t_i}
\le2^{-4}U_{i-1}^{-3}\le2^{-3\ell}.
\]
After $s$ successful stages the output is a width-$k$ CNF or DNF. A CNF has correlation at most $\kappa_k$. A DNF equals $1-H$ with $H$ a width-$k$ CNF, and therefore has correlation at most $2\kappa_k$, because the constant-one test is also included. The restricted target need not be balanced. Use the Boolean bound one on failure and enlarge the successful expectation to the unconditional expectation over $\phi$. This proves Eqn.~\eqref{eq:ordinary-switch}.

\subsection{Common partial decision trees}
Fix a positive integer $D_0$. Define
\begin{equation}
R_0=0,\qquad D_i=R_{i-1}+D_0,\qquad
R_i=R_{i-1}+D_i=(2^i-1)D_0.
\label{eq:tree-budgets}
\end{equation}
After stage $i-1$, maintain a restriction tree of depth at most $R_{i-1}$. Every active leaf has depth reduced by $i-1$, size at most $U_{i-1}$, and bottom fan-in at most $k_{i-1}$. Frozen leaves satisfy the same bounds.

Condition on earlier randomness. An old leaf has a fixed path restriction $\tau$. On coordinates outside $\operatorname{dom}\tau$, the next global random restriction is an independent $p_i$-random restriction. Apply the multi-switching lemma to the at most $U_{i-1}$ second-layer functions of that cofactor, requesting depth $D_i$ and residual depth at most $t_i$. Require this to succeed for every old leaf, even those that the new random restriction will make incompatible. A union bound over at most $2^{R_{i-1}}$ leaves gives conditional failure at most
\begin{equation}
2^{R_{i-1}}U_{i-1}(24p_i k_{i-1})^{D_i}
\le2^{R_{i-1}}U_{i-1}4^{-D_i}
=U_{i-1}2^{-R_{i-1}-2D_0}.
\label{eq:common-failure}
\end{equation}
Frozen leaves need no call.

Restrict the old tree by the global random restriction and delete incompatible paths. At each surviving active leaf splice in its new common tree, which queries only coordinates outside that old path. The resulting tree has depth at most $R_i$. Switching the residual depth-two functions to their opposite forms reduces circuit depth by one at each new leaf. The size and bottom fan-in bounds follow from Eqn.~\eqref{eq:budgets}. This proves the induction.

Crucially, the $2^{R_{i-1}}$ factor is charged only to failure probability. It does not multiply the size of an individual leaf circuit or its bottom fan-in. We use a product distribution only for the unconditional composition of the global random restrictions in Eqn.~\eqref{eq:global-survival}.

Summing Eqn.~\eqref{eq:common-failure} over $s$ stages bounds total failure by $s2^{A\ell-2D_0}$. On success, \Cref{lem:guards} bounds the final correlation by $2^{R_s+1}\kappa_k(g_{V,\phi})$, even though its atoms depend on the full history. Enlarge to the unconditional expectation to obtain Eqn.~\eqref{eq:common-switch}.

\section{Uniform parameter checks for average-case hardness}
\label{app:parameters}
We give explicit depth constants to substantiate the uniform size range, rather than hide endpoint conditions in asymptotic notation. Let
\begin{align*}
s&=d-2,& K&=9\cdot6^d,& A&=6^d,\\
Q&=(96K)^{-s},& a&=\frac{Q}{2^{20}K},& J&=2^{d+10},\\
M_d&=4JA,& \zeta&=\frac1{16M_d},&
c_*&=(a/128)^{1/(d-1)}.
\end{align*}
The symbol $M_d$ here is a depth constant, not an input dimension. For all sufficiently large $N$, the deterministic bounds in Eqn.~\eqref{eq:budgets} satisfy
\begin{equation}
\ell\le k_i\le K\ell,\qquad
Q\ell^{-s}\le q\le(96\ell)^{-s},\qquad
\ell\le k\le K\ell.
\label{eq:uniform-budgets}
\end{equation}
Assume Eqn.~\eqref{eq:average-range}. For any subcube of codimension less than $N/4$, set
\[
b=\floor{\frac{\log k}{2048}},\quad
p=\frac b{64k},\quad m=(\dim V)qp,\quad
\mathcal R=\frac{N\log\ell}{\ell^{d-1}}.
\]
Uniformly in this range, eventually
\begin{equation}
\frac{\log\ell}{4096}\le b\le\frac{\log\ell}{1024},\qquad
3a\mathcal R\le m\le\frac{\mathcal R}{2^{16}96^s}.
\label{eq:m-bounds}
\end{equation}
The lower bound uses $\dim V\ge3N/4$ and Eqn.~\eqref{eq:uniform-budgets}; the upper bound uses $\dim V\le N$ and $k\ge\ell$. In particular $k\ge b\ge2$ eventually, and the upper-size constant ensures $m\ge128\log\ell$.

For the list-length bound in Eqn.~\eqref{eq:inverse-parameters}, $2^{512b}\le k^{1/4}$ gives
\begin{equation}
L=O_d(\ell^{1/4}\log\ell\log N),\qquad
\log(8L)\le\tfrac14\log\ell+\log\log\ell+\log\log N+O_d(1)
\le2\log\ell.
\label{eq:L-bounds}
\end{equation}
The last estimate uses $\ell\ge\log N$. Hence $m\ge16\log(8L)$ and $m\ge2b$. Also
\begin{align*}
\frac{mL}{N}
&=O_d\left(\ell^{-(d-5/4)}(\log\ell)^2\log N\right)\\
&\le O_d\left(\ell^{-(d-9/4)}(\log\ell)^2\right)=o(1).
\end{align*}
This is uniform down to polynomial-size circuits: at $\ell=\log N$ and $d=3$ the upper bound is $O_d((\log\log N)^2/(\log N)^{3/4})$. Thus $mL\le N/64$ eventually. Furthermore $m=o(N)$ uniformly, and $L$ is polynomially bounded in $N$.

All terms in $\Delta$ in Eqn.~\eqref{eq:inverse-result} are eventually bounded by a constant multiple of $\gamma=2^{-3m/64}$: the two dimension-tail contributions and the paired term have stronger exponents in $m$, and $128L2^{-3N/16}$ is exponentially smaller since $m=o(N)$. In particular Eqn.~\eqref{eq:inverse-simplified} follows.

If $m<M_d\ell$, Eqns.~\eqref{eq:ordinary-switch} and~\eqref{eq:inverse-simplified} give
\[
|\E_V[fC]|\le s2^{-3m/M_d}+10\cdot2^{-3m/64}.
\]
If $m\ge M_d\ell=4JA\ell$, set $D_0=\floor{m/J}$. For large $N$, $m\ge2J$ and
\[
A\ell-2D_0\le-m/J,\qquad R_s\le2^{d-2}m/J=m/4096.
\]
The multi-switching estimate gives
\[
|\E_V[fC]|\le s2^{-m/J}+10\cdot2^{-(3/64-1/4096)m}.
\]
Since $m\to\infty$ uniformly and $\zeta=1/(16M_d)$ is smaller than the displayed decay constants, both cases are at most $2^{-\zeta m}$ eventually. Using $m\ge a\mathcal R$ proves \Cref{prop:narrow} with $\theta_d=\zeta a$.

Finally, $\mathcal R/N=\log\ell/\ell^{d-1}\to0$ uniformly. For large $\ell$ the function $\log\ell/\ell^{d-1}$ decreases, so its minimum over the range in Eqn.~\eqref{eq:average-range} is at the upper endpoint, where $\mathcal R\ge c_*^{-(d-1)}\log\ell\to\infty$. This justifies absorbing fixed factors and the cofactor-recursion remainder into a smaller exponent constant.

\section{Uniform fields, general circuits, and padding}
\label{app:uniform}
\subsection{A deterministic field tower}
Start with $E_0=\F_2$ and $\gamma_0=1$. Recursively define
\[
E_j=E_{j-1}[\theta_j]/(\theta_j^2+\theta_j+\gamma_{j-1}),\qquad
\gamma_j=\gamma_{j-1}\theta_j.
\]
Inductively $\gamma_{j-1}$ has absolute trace one. In a binary field, $z\mapsto z^2+z$ has kernel $\{0,1\}$ and its image is the trace-zero subspace. Indeed trace vanishes on its image, and both spaces have codimension one. Therefore the displayed quadratic is irreducible. Its roots differ by one, so $\Tr_{E_j/E_{j-1}}(\theta_j)=1$. It follows that
\[
\Tr_{E_j/\F_2}(\gamma_j)=\Tr_{E_{j-1}/\F_2}(\gamma_{j-1})=1.
\]
Thus $E_j$ has binary dimension $N=2^j$. The recursive pair basis is deterministic. Multiplication uses at most five multiplications in the previous field and a linear number of bit additions, so its complexity is polynomial in $N$. Inversion by $x^{2^N-2}$ for $x\ne0$, and trace by $N$ Frobenius iterates, are also polynomial time. No search for irreducible polynomials is needed. The finitely many small input lengths can be defined by the same inverse-trace rule.

\subsection{Layering a circuit without increasing its depth}
Let a general circuit have size $S_0\ge N$ and depth at most $d$. First bypass paths of consecutive same-type gates by connecting each parent directly to its same-type descendants' frontier inputs. Retain a shared gate when it is still used by an opposite-type parent. This preserves the Boolean function without adding gates or increasing depth. Every remaining internal wire joins opposite gate types.

All gate-to-output path lengths from a given gate have parity determined by its type. Assign that gate its longest such path level. An internal wire skips a positive odd number of levels. Fill the skipped levels with alternating unary gates; similarly pad literal-to-gate wires. There are $O((S_0+N)^2)$ distinct wires and $d$ is fixed, so this adds $O_d(S_0^2)$ gates and does not increase depth. Unused layers, constants, and unary gates may be retained. This representation supports the fixed layer labels used in cofactor recursion.

For sufficiently large $S_0$, the deterministic budget $S=S_0^3$ dominates the normalized size. With $\ell=3\log S_0$, the previous estimates apply for
\[
N\le S_0\le2^{(c_*/3)N^{1/(d-1)}}.
\]
Changing $\log S_0$ to $3\log S_0$ alters $N\log\log S_0/(\log S_0)^{d-1}$ by depth-dependent constant factors. Eqn.~\eqref{eq:boolean-sign} absorbs the factor two in passing to sign correlation.

\subsection{A single family at all lengths}
For arbitrary $n$, let $N$ be the largest power of two at most $n$, and define $F_n$ to evaluate $I_N$ on its first $N$ input bits. Ignore the remaining bits. Then $N\le n<2N$, and this is one uniform polynomial-time family, independent of $d,S$.

Fixing the ignored bits of an $n$-input circuit leaves a circuit on $N$ bits of no larger size or depth. The tower-length bound holds for every such fixing, so average and use the triangle inequality. Since $S\ge n\ge N$ and $N>n/2$, reducing the size constant by a factor $2^{1/(d-1)}$ and reducing the exponent constant by an absolute factor proves \Cref{thm:average}. For worst-case hardness, any exact circuit restricts to an exact circuit for $I_N$; moreover $N\log N=\Theta(n\log n)$. This proves the padded part of \Cref{thm:worst}. No pseudorandomness under translations in ignored directions is asserted or needed.

\section{The two parameterized worst-case targets}
\label{app:other-targets}
\subsection{The trilinear function}
Write $g_z=(-1)^{T_a(\cdot,\cdot,z)}$. We use the following two results of Chukhin~\cite[Lemmas 4.2 and 4.3]{Chukhin26}. If $T_y\subseteq\{0,\ldots,a-1\}$ includes each index independently with probability $m/a$, where $m\le a^{3/5}$ and $u\le m^{1/4}$, then except with probability $a^{-\Omega(m)}$,
\begin{equation}
|B+T_y|\ge|B|(|T_y|-m/4)
\quad\text{for all }B\subseteq\{0,\ldots,a-1\},\ |B|\le u.
\label{eq:sumset}
\end{equation}
If a fixed data restriction has $|T_x|\ge2u$, $u$ even, and $|B+T_y|\ge|B||T_y|/2$ for those $B$, then for every fixed Boolean indicator $J$ on the cube of free variables,
\begin{equation}
\E_z\abs{\E[g_{z,\sigma}J]}^u
\le2^{-u|T_x|/4}+2^{-u|T_y|/4}.
\label{eq:trilinear-moment}
\end{equation}
The indicator can be arbitrary; a clause description is not needed for this moment input.

\begin{proposition}[A hard trilinear slice]
\label{prop:trilinear}
Let $k\ge b\ge2$, $0<q\le1$, $p=b/(64k)$, and $m=qpa$. Let $u$ be the largest even integer at most $m^{1/4}$ and set
\[
L_* =\ceil{2^{2048b}\bigl[b\ln(8m+1)+1\bigr]}.
\]
If $m$ is sufficiently large, $m\le a^{3/5}$, and $L_*\le u/128$, then some fixed $z$ satisfies
\[
\E_\phi\kappa_k(g_{z,\phi})\le5\cdot2^{-m/32},
\]
where $\phi$ is a $q$-restriction of the $2a$ data bits.
\end{proposition}
\begin{proof}
Put $\eps=2^{-m/32}$ and draw the fresh inner $p$-restriction. The numbers of free variables in the two blocks are in $[m/2,2m]$ except with probability at most $2\eps$ eventually. Eqn.~\eqref{eq:sumset} fails with probability at most $\eps$. Outside these events, the total number of free variables $v$ lies in $[m,4m]$, the sumset hypothesis of Eqn.~\eqref{eq:trilinear-moment} holds, and $|T_x|\ge2u$.

Markov's inequality at threshold $2^{-m/16}$ bounds the failure probability for any fixed list test by $2^{1-um/16}$. Union over fewer than $2^{vL_*+1}$ tests to get
\[
2^{4mL_*+2-um/16}\le2^{2-um/32}\le\eps
\]
eventually. On success, \Cref{thm:partition} with $C=2048$ gives for every width-$b$ CNF $H$
\[
|\E[g_{z,\phi\rho}H]|\le2^{v/128}2^{-m/16}\le\eps.
\]
The joint failure probability over $z,\phi,\rho$ is at most $4\eps$. Averaging fixes one $z$ with failure probability at most $4\eps$. \Cref{lem:transfer} then gives $5\eps$.
\end{proof}

To finish the lower bound, fix $d\ge3$ and let
\[
\eta=\frac1{32\cdot2048(d-1)},\quad
b=\floor{\eta\log a},\quad
\ell=c_d(a\log a)^{1/(d-1)},\quad S=2^\ell.
\]
Use the worst-case depth-reduction parameters $k,q$ from \Cref{sec:worst-proof}. For small fixed $c_d>0$, $k\ge b$ eventually, and
\[
m=\Theta_d\bigl(\eta c_d^{-(d-2)}(a\log a)^{1/(d-1)}\bigr).
\]
Then $m\le a^{3/5}$ and
\[
L_*\le a^{1/[32(d-1)]}O_d((\log a)^2)=o(m^{1/4}).
\]
\Cref{prop:trilinear} applies. The constant-one CNF and restriction averaging give $|\E g_z|\le5\cdot2^{-m/32}$, so both level sets of the selected slice have density at least $1/3$. \Cref{lem:cover} would imply
\[
\frac13\le\frac14+5S^{A_d}2^{-m/32}.
\]
Since $m/\ell=\Theta_d(\eta c_d^{-(d-1)})$, sufficiently small $c_d$ gives a contradiction. The total input length is $4a-1$, proving the trilinear part of \Cref{thm:worst}.

\subsection{The original polynomial evaluator}
\label{sec:evaluator}
We spell out the family to avoid changing its input accounting or hiding its parameter inputs. At total input length $n$, let $a=\floor{n/5}$, $r=\ceil{a^{2/3}}$, and let $t$ be the largest nonnegative even integer at most $a^{1/6}$. Reject if $a<64$ or if the following blocks do not fit. Otherwise parse:
\begin{itemize}
\item $a$ data bits $x_1,\ldots,x_a$ and $a+r-1$ hash bits $u_1,\ldots,u_{a+r-1}$;
\item $r$ bits specifying the monic modulus $P(Z)=Z^r+\sum_{i<r}p_iZ^i$;
\item $tr$ coefficient bits specifying $\beta_0,\ldots,\beta_{t-1}$ in $R_P=\F_2[Z]/(P)$.
\end{itemize}
Their total is $2a+(t+2)r-1\le n$ for all sufficiently large lengths. Ignore unused bits. Form
\[
h_i(x)=\sum_{s=1}^a u_{i+s-1}x_s\pmod2\quad(1\le i\le r),\qquad
h(x)=\sum_{i=1}^r h_i(x)Z^{i-1}\in R_P.
\]
Let $\lambda$ select the degree-zero coefficient. The evaluator accepts exactly when
\[
\lambda\left(\sum_{j<t}\beta_j h(x)^j\right)=0.
\]
This is the original family of~\cite{OpenAI26}. Ring operations are defined for every monic input modulus, whether reducible or not, and the algorithm runs in polynomial time. No irreducibility search is part of computing the function.

For hardness analysis choose one irreducible modulus of degree $r$. For every nonzero difference $w$ of data inputs, the hash $h(w)$ is uniform over $\F_2^r$ when $u$ is uniform. To see this, let $s$ be the largest coordinate with $w_s=1$. Row $i$ has the fresh pivot $u_{i+s-1}$, with all other indices smaller, so the $r$ rows are independent linear forms in $u$. A $v$-dimensional coordinate subcube has at most $2^v-1$ nonzero differences; the hash fails to be injective on it with probability at most $2^{v-r}$.

Conditioned on an injective hash, evaluation at any at most $t$ distinct hashes maps uniform coefficients to independent uniform field values. This follows from Lagrange interpolation with polynomials of degree at most $t-1$. Applying the nonzero coordinate functional $\lambda$ gives unbiased $t$-wise independent signs $g$ on the cube of free variables. For any fixed indicator $J$, expand the even $t$th moment of $Y=\sum_z g(z)J(z)$. A tuple with a singly occurring point contributes zero. The remaining tuples have at most $t/2$ distinct points and are covered by at most $2^{vt/2}t^t$ descriptions. Therefore
\begin{equation}
\Prb_\beta[|\E[gJ]|>2^{-v/4}]\le t^t2^{-vt/4}.
\label{eq:evaluator-tail}
\end{equation}
This argument derives the moment directly for arbitrary indicators; it does not invoke a source proposition outside its fixed-width quantifiers.

For an expected number of free variables $m$, set
\[
L_* =\ceil{2^{512b}\bigl[b\ln(4m+1)+1\bigr]}.
\]
For $v\in[m/2,2m]$, union over fewer than $2^{vL_*+1}$ tests. By Eqn.~\eqref{eq:evaluator-tail}, failure is at most $2^{-vt/8}$ whenever
\[
L_*/t\le1/32,\qquad
\frac{\log t}{m/2}\le1/32,\qquad
\frac1{(m/2)t}\le1/32.
\]
Indeed these bounds leave the total positive exponent at most $3vt/32<vt/8$. On success, the $C=512$ point-list partition gives $\kappa_b(g)\le2^{v/32}2^{-v/4}\le2^{-v/8}$. If $r\ge(2+1/16)m$, the hash error is at most $2^{-m/16}$. Each dimension tail is also at most $2^{-m/16}$. Thus joint failure is at most $4\cdot2^{-m/16}$ eventually. Averaging fixes all non-data inputs before selecting any circuit or test, and \Cref{lem:transfer} gives a slice with
\[
\E_\phi\kappa_k(g_\phi)\le5\cdot2^{-m/16}.
\]

Finally choose $b=\floor{\log a/(64\cdot512)}$, $\ell=c_d(a\log a)^{1/(d-1)}$, and the worst-case parameters $k,q$, $p=b/(64k)$. Then
\[
m=\Theta_d\bigl(c_d^{-(d-2)}(a\log a)^{1/(d-1)}\bigr),\quad
L_*\le a^{1/64}O((\log a)^2)=o(t),\quad
\log t=o(m),\quad r/m\to\infty.
\]
All preceding conditions hold. The constant-one test gives output densities at least $1/3$. \Cref{lem:cover} contradicts exact computation by size $2^\ell$ when $c_d$ is sufficiently small. Since $a=\floor{n/5}$, this proves the evaluator part of \Cref{thm:worst} in total input length, with its original parameters unchanged.

\section{Further consequences of the decomposition}
\label{app:refinements}
\subsection{Bounded-width CNFs and monochromatic sets for inverse trace}
\begin{corollary}
\label{cor:wide}
For all sufficiently large tower lengths $N$, set $C_0=512$ and let
\[
(\log N)^2\le k\le\frac{N}{4096C_0},\qquad
b=\floor{\frac{\log k}{4C_0}}.
\]
For $f=(-1)^{I_N}$ and every width-$k$ CNF $H$,
\[
|\E[fH]|\le5\cdot2^{-3Nb/(4096k)}.
\]
If $H\ne\varnothing$ is monochromatic for $I_N$, then
\[
-\log\Prb[H=1]\ge\frac{3Nb}{4096k}-\log5
=\Omega(N\log k/k).
\]
\end{corollary}
\begin{proof}
Use \Cref{prop:inverse} with $q=1$ and $V=E$, so $m=Nb/(64k)$. Its list-length bound obeys
\[
L=O(k^{1/4}\log k\log N),\qquad
bL=O(k^{1/4}(\log k)^2\log N)=o(k)
\]
uniformly in the stated range. This gives $mL\le N/64$ eventually. For $k\ge\sqrt N$, the upper endpoint gives $m\ge64C_0b$, while $\log(8L)\le2C_0b$ eventually. For $k<\sqrt N$, $m\ge\sqrt N b/64\gg\log L$. Thus the remaining hypotheses hold. The terms of $\Delta$ are bounded by four times $2^{-3m/64}$ eventually. For a monochromatic CNF $H$, we have $|\E[fH]|=\Prb[H=1]$.
\end{proof}

\subsection{The intrinsic list-length requirement}
\begin{proposition}
\label{prop:list-lower}
Let $1\le b\le m$. A nonempty $J=\cl_b(A)$ with $|A|\le L$ has $|J|\le L^{m/b}$. If $H\ne\varnothing$ has an exact signed representation by such tests, then Eqn.~\eqref{eq:list-lower} holds.
\end{proposition}
\begin{proof}
For $X$ uniform on $J$, each projection $X_U$, $|U|=b$, has entropy at most $\log L$. Expand every $H(X_U)$ in coordinate order. Conditioning on only $U\cap[i-1]$ cannot reduce the conditional entropy below $H(X_i\mid X_{[i-1]})$. Each coordinate appears in $\binom{m-1}{b-1}$ subsets, whence
\[
\log|J|=H(X)\le\frac1{\binom{m-1}{b-1}}\sum_{|U|=b}H(X_U)
\le\frac mb\log L.
\]
For a signed representation, sum its pointwise absolute upper bound over the cube:
\[
|H|\le\sum_i|c_i|\,|\cl_b(A_i)|\le L^{m/b}\sum_i|c_i|.
\]
\end{proof}

\section{Comparison with the Fourier-tail bound}
\label{app:fourier}
For a binary character $\chi_\xi(x)=(-1)^{\langle\xi,x\rangle}$, nondegeneracy of the trace pairing supplies $a\in E$ with $\chi_\xi(x)=\psi(ax)$. The case $h=1$ of \Cref{lem:shifts} gives
\[
\max_\xi|\widehat f(\xi)|\le8\cdot2^{-N/2}.
\]
For a sign-valued depth-$d$ circuit $G$ of size $S$, Tal's Fourier-tail theorem~\cite[Theorem 1]{Tal17} gives, for a depth constant $A_d$,
\[
\sum_{|U|\ge k}\widehat G(U)^2
\le2\cdot2^{-k/(A_d(\log S)^{d-1})}.
\]
Take $k=\floor{N/100}$. The Fourier mass of $f$ below $k$ is at most
\[
64\cdot2^{-N}\sum_{j<k}\binom Nj
\le2^{-(1-h_2(1/100))N+O(1)}.
\]
Split Parseval's identity at $k$ and apply Cauchy--Schwarz to each part. Using the fact that each sign function has total Fourier mass one gives
\[
|\E[fG]|\le2^{-\Omega(N)}+\sqrt2\,2^{-k/(2A_d(\log S)^{d-1})}.
\]
This is $2^{-\Omega_d(N/(\log S)^{d-1})}$ when that exponent tends to infinity, with the displayed estimate valid also outside that regime. Padding works as in \Cref{app:uniform}. The higher shift moments, point-list decomposition, and guarded-CNF argument supply the additional $\log\log S$ factor in \Cref{thm:average}.

\end{document}